\documentclass[flushrt]{aastex}
\usepackage{emulateapj5}

\newcommand{\hi}{H\,{\sc i}}

\shorttitle{Radio-Selected \hi\ Absorption}
\shortauthors{Darling et al.}

\begin{document}


\title{Detection of 21 Centimeter \hi\ Absorption at $z=0.78$ in a 
Survey of Radio Continuum Sources}

%


\author{Jeremy Darling}
\affil{Carnegie Observatories, 813 Santa Barbara Street, 
    Pasadena, CA 91101}
\email{darling@ociw.edu}
\author{Riccardo Giovanelli \& Martha P. Haynes}
\affil{Department of Astronomy and Space Sciences, Cornell University, 
    Ithaca, NY  14853}
\email{riccardo@astro.cornell.edu, haynes@astro.cornell.edu}
\author{Alberto D. Bolatto \& Geoffrey C. Bower}
\affil{Department of Astronomy, University of California at Berkeley, 
	Berkeley, CA  94720-3411}
\email{bolatto@astro.berkeley.edu, gbower@astro.berkeley.edu}

\begin{abstract}
We report the detection of a deep broad \ion{H}{1} 
21 cm absorption system at $z=0.78$
toward the radio source [HB89] 2351+456 (4C+45.51) at $z=1.992$.  
The \ion{H}{1} absorption was identified in a blind spectral line survey 
conducted at the Green Bank Telescope
spanning $0.63<z<1.10$ toward a large sample of radio continuum sources.  
The \ion{H}{1} column density
is $N_{HI} = 2.35 \times 10^{19} (T_s/f)$ cm$^{-2}$, where $T_s$ is the 
spin temperature and $f$ is the continuum covering factor of the absorbing
gas.  For $T_s/f > 8.5$ K, 
this system is by definition a damped Ly$\alpha$ absorption system 
($N_{HI} \geq 2\times10^{20}$ cm$^{-2}$).  The line is unusually broad, 
with a FWHM of 53 km s$^{-1}$ and a full span of 163 km s$^{-1}$, 
suggesting a physically extended \ion{H}{1} gas structure.  
Radio surveys
identify damped Ly$\alpha$ systems in a manner 
that bypasses many of the selection effects present in optical/UV surveys,
including dust extinction and the atmospheric cutoff for $z<1.65$.
The smooth broad profile of this \ion{H}{1} 21 cm absorption system is similar
to the $z=0.89$ \ion{H}{1} absorption toward PKS 1830$-$211, which suggests
that the absorber toward [HB89] 2351+456 is also a gravitational lens 
and a molecular absorption system.  However, 
very long baseline interferometry
and {\it Hubble Space Telescope}
observations show little evidence for gravitational lensing, and
BIMA millimeter observations show no HCO$^+$ ($1\rightarrow2$)
or HCN ($1\rightarrow2$) absorption down to $\tau=0.15$ ($3\sigma$)
in 5 km s$^{-1}$ channels.  Although this radio 
damped Ly$\alpha$ selection technique would include dusty, molecule-rich 
systems, [HB89] 2351+456 appears to be a ``vanilla'' \ion{H}{1} 21 cm 
absorber.  
\end{abstract}


\section{Introduction}
Identification of damped Ly$\alpha$ (DLA) systems, quasar absorption
line systems with neutral hydrogen column density 
$N_{HI} \geq 2\times10^{20}$ cm$^{-2}$, has historically been 
an optical/UV pursuit.  DLA catalogs suffer from the 
selection effects associated with an optically selected sample of redshifted
Ly$\alpha$ lines, including the atmospheric cutoff at $z<1.65$, the 
requirement of a strong UV continuum, and the potential loss of dusty
DLA systems.  To overcome the UV atmospheric cutoff, low ionization 
absorption lines such as 
\ion{Mg}{2} $\lambda$2796, 2803 and \ion{Fe}{2}
$\lambda$2600 provide proxies for DLA systems that can subsequently
be observed in the Ly$\alpha$ line from space or in the \ion{H}{1} 21 cm
line from the ground \citep{rao00,lane00}.  The \ion{H}{1} 21 cm line, unlike
Ly$\alpha$, does not saturate, is unaffected by dust, and can occasionally
offer multiple lines of sight through the absorption system.  
\ion{H}{1} 21 cm absorption is a prerequisite for molecular absorption, so 
a dust-independent DLA selection method maximizes the likelihood of adding
to the very small sample of extragalactic molecular absorption systems.  
These systems provide insight into the physical state of the cold 
interstellar medium in 
galaxies and can constrain the cosmic evolution of physical constants 
\citep[e.g.,][]{wik97,dar04,kan04}.  
While there does not appear to be a dominant undetected population of high
column density DLA systems compared to the DLA population overall,
optically selected samples may exclude a large fraction of the dusty
absorption line systems that do exist \citep{car98,ell01},
thereby significantly 
diminishing the likelihood of identifying molecular absorption systems.

Traditional
approaches to \ion{H}{1} 21 cm observations of DLA systems have focused on 
identifying the radio line toward optically identified systems of known 
redshift \citep[e.g.,][]{car96}, although there have also been 
``blind'' narrow 
instantaneous bandwidth surveys for intervening \ion{H}{1} absorption and
surveys for intrinsic absorption in quasars \citep[e.g.,][]{bro73,car98}.
New radio astronomy facilities equipped with broadband receivers and 
high resolution spectrometers now offer the opportunity to conduct 
line searches
for DLA systems in a manner similar to traditional optical/UV DLA surveys, 
spanning large redshift intervals with high spectral resolution.  
For example, the Green Bank Telescope\footnote{The 
National Radio Astronomy Observatory is a facility of the 
National Science Foundation operated under cooperative agreement by 
Associated Universities, Inc.} (GBT)
offers continuous frequency coverage from 290 to 1420 MHz ($z=0$--3.9
in \ion{H}{1}), large instantaneous redshift coverage, and high
velocity resolution (e.g., $\Delta z = 0.5$ and $\Delta v = 5$ km s$^{-1}$
at 800 MHz).

We have conducted a DLA survey toward strong radio sources at the 
GBT at 800 MHz
spanning the redshift range $z=0.63$--1.10 with 5 km s$^{-1}$ resolution
that is sensitive to {\it all} DLA systems.  We select sources 
with predicted 780 MHz flux density $\geq1.7$ Jy, as interpolated from 1.4 GHz 
\citet{whi92} and 365 MHz \citet{dou96} measurements.  
This is not a perfect radio survey:  we also require sources to have
$z>1.06$, which biases the sample
toward systems with optical identifications and measurable lines.  

We report the first detection of the Green Bank DLA survey, 
a deep broad \ion{H}{1} absorption 
system at $z=0.78$ toward the radio source [HB89] 2351+456 (hereafter 
2351+456, but also known as
4C+45.51, B3 2351+456, NVSS J235421+455304, and WMAP 074, among others).  
To investigate the nature of this absorption system, we present 
BIMA millimeter observations of HCO$^+$ and HCN.
We also reexamine 
published optical spectra and infrared photometry and find that the \ion{H}{1}
absorption system provides new interpretations of observed properties of 
2351+456.  
This Letter assumes the concordance cosmology:  
$H_\circ = 71$ km s$^{-1}$ Mpc$^{-1}$, $\Omega_M = 0.27$, and 
$\Omega_\Lambda = 0.73$.

\section{Observations}

\subsection{Green Bank Telescope}

We observed 2351+456 at 773.75 MHz spanning 
192.5 MHz at the GBT on 2004 January 26.
We observed four 50 MHz IFs with 2.5 MHz overlap in two linear
polarizations, spanning the range 677.5--870 MHz or $z_{HI} = 0.63$--1.10.  
Each bandpass was broken into 8192 6.1 kHz channels, but adjacent channels
in the autocorrelation spectrometer are not independent and the 
effective spectral resolution is 
12.2 kHz or 4.7 km s$^{-1}$ at the band center ($z=0.84$).  Fast sampling 
(1.5 s data records) was used to facilitate radio
frequency interference (RFI) excision.
Observations of 2351+456 include a single 5 minute position-switched
observation and a 25 minute on-source scan.  

An \ion{H}{1} absorption line was identified at 798.2 MHz ($z=0.78$) in 
individual 1.5 s records during observations.  
We find no evidence for RFI contamination of the detected 21 cm line, and
we find no evidence that the line is an RFI signal in the off-source 
scan or an out-of-band harmonic; the HI absorption line is clearly visible
in individual records, and this line is not seen in any other program source.

\begin{figure}
\plotone{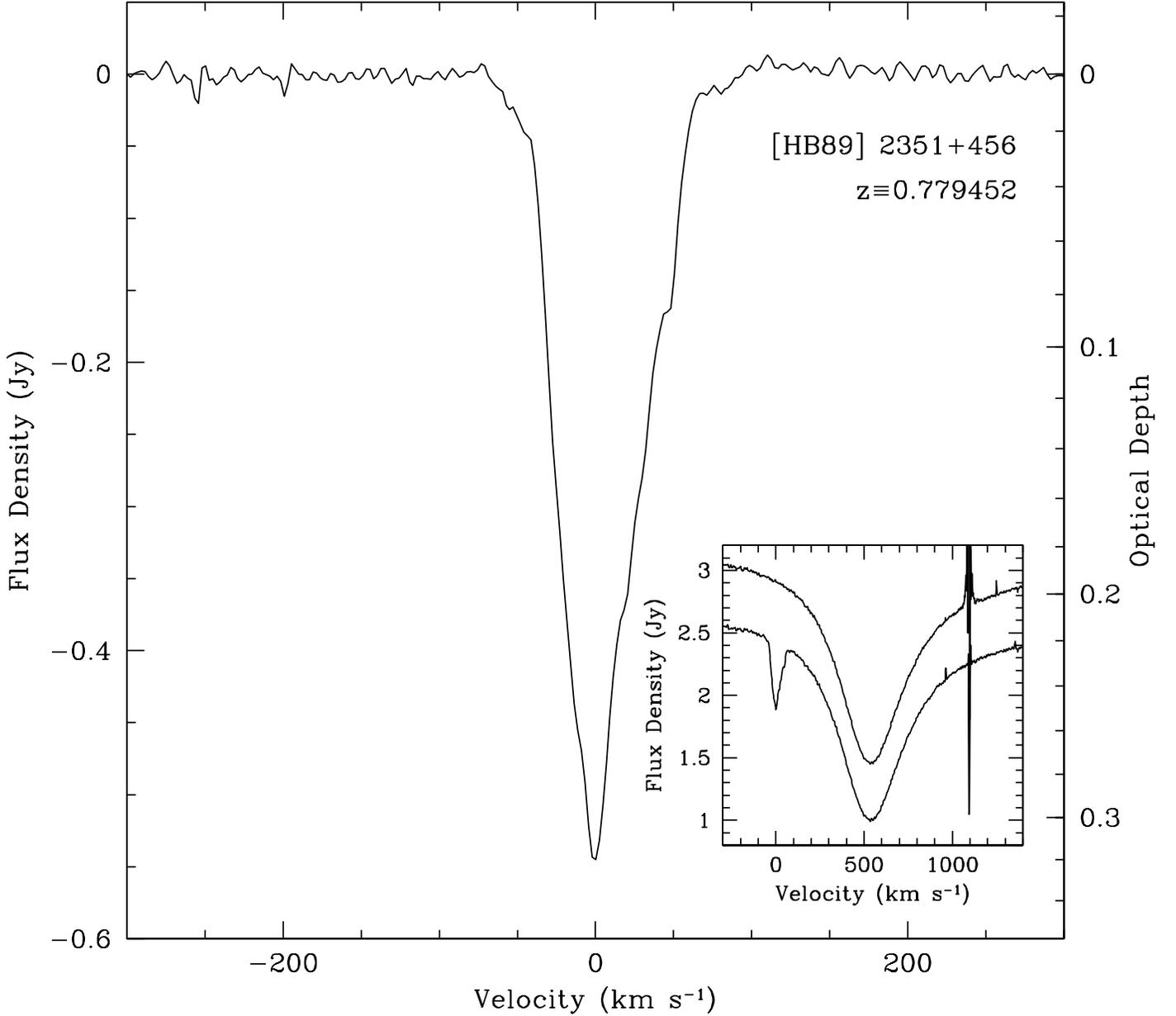}
\caption{The \ion{H}{1} absorption spectrum toward 2351+456.  
Zero velocity refers to barycentric redshift $z=0.779452$, 
and the velocity scale is in the rest frame of the absorption system.  
Zero flux density and zero optical depth correspond to the continuum level 
of 2351+456, assumed to be 1.99 Jy.
{\it Inset:}  Position-switched spectra of 2351+456 ({\it bottom})
and [HB89] 2338+042 ({\it top}) 
centered on the 800 MHz receiver feed resonance
at 796.8 MHz.  The spectrum of [HB89] 2338+042 has been shifted by +0.5 Jy.
[HB89] 2338+042 provides a template for the feed resonance and general
bandpass subtraction of the 2351+456 spectrum.  The narrow features
on the red side of the resonance are RFI.
\label{fig:spec}}
\end{figure}

Records were individually calibrated and bandpasses flattened using the 
winking calibration diode.  The 25 minute on-source scan was calibrated
in a position-switched mode using the 5 minute off-source scan at the cost
of added noise.  
A 25 minute scan on [HB89] 2338+042 observed during the same session as
2351+456 was calibrated in an identical manner as
a template for subtraction of the feed resonance from the 2351+456
spectrum (Figure \ref{fig:spec}, inset).  A least-squares fit of the template
spectrum to the 2351+456 spectrum in line- and RFI- free regions
indicated that a small shift (0.073 Jy) and scaling (0.975) was adequate 
to fit the resonance in 2351+456.  We subtracted the shifted and scaled 
template from the 2351+456 spectrum to obtain a resonance and baseline
subtracted spectrum.  
A polynomial baseline was subtracted from a
fit of 5 MHz centered on the \ion{H}{1} 
line to remove bandpass ripples of order
2 MHz (the total span plotted in Figure \ref{fig:spec} is 1.6 MHz).
We reached a typical
rms noise of 4 mJy in 6.1 kHz (2.3 km s$^{-1}$) channels.  
All data reduction was performed in AIPS++.\footnote{The AIPS++ 
(Astronomical Information Processing System) 
is freely available for use under 
the Gnu Public License. Further information may be obtained from 
http://aips2.nrao.edu}

\subsection{BIMA HCO$^+$ and HCN Observations}

We observed 2351+456 in the 
HCO$^+$ $J=1\rightarrow2$ (178.3751 GHz) and 
HCN $J=1\rightarrow2$ (177.2611 GHz) lines redshifted to 
100.2344 and 99.60678 GHz, respectively (assuming $z=0.7796$), 
at the Berkeley-Illinois-Maryland
Association (BIMA; \citealt{Wea96}) array at Hat Creek on 2004 April 9 and 18.
We used correlator mode 6, which provides four independent spectral
windows. Two of them are 100~MHz wide with a spectral resolution
$\sim1.56$ MHz ($\sim5$ km~s$^{-1}$), and 
two are 200~MHz wide with a
resolution $\sim3.12$ MHz ($\sim10$ km~s$^{-1}$).  
The two high-resolution
windows were partially overlapped and centered on each expected
absorption feature, covering a velocity range of $\Delta v\sim\pm200$
km~s$^{-1}$ around it. The two low-resolution windows extended this
velocity coverage to $\Delta v\sim\pm750$ km~s$^{-1}$.

\begin{figure*}
\plotone{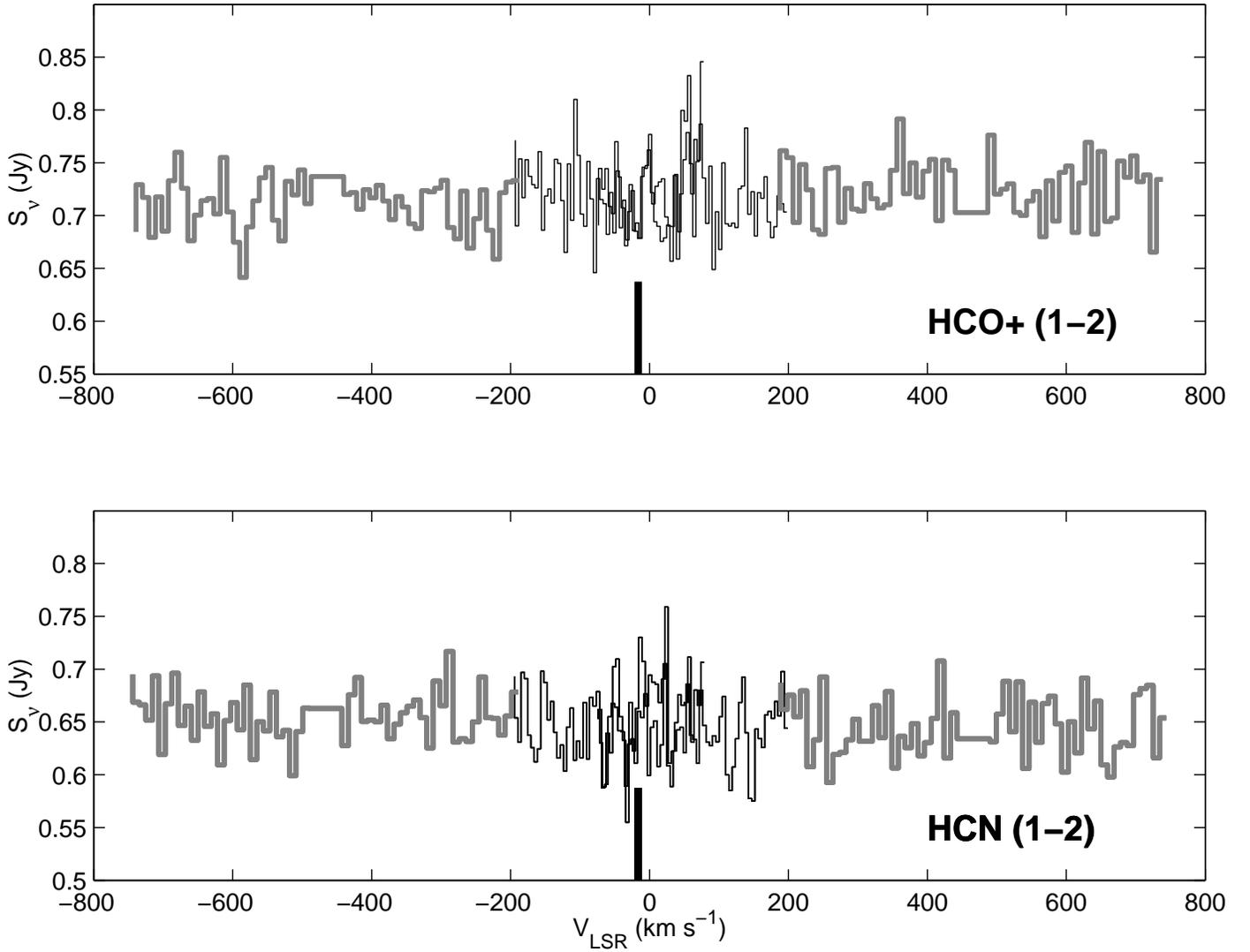}
\caption{Millimeter-wave spectra of 2351+456 obtained at BIMA for
the redshifted transitions of HCO$^+$ ($1\rightarrow2$) and HCN 
($1\rightarrow2$) toward 2351+456.
The thin trace corresponds to the spectral region observed at 
high resolution, while the thick gray trace corresponds to the low
resolution correlator windows. The thick black line at the bottom
illustrates the expected $V_{LSR}$ of the molecular absorption for
$z=0.779452$, taking the heliocentric to LSR frame corrections
into account.
\label{mmspectra}}
\end{figure*}

Observations of a passband
calibrator were interleaved (a few minutes every hour) with the
science observations and used to remove passband features. During the
first track on April 9, we used QSO 0102+584, which was
considerably weaker ($S_\nu\sim1.5$ Jy) than predicted by the flux
database and introduced some noise into the passband correction.  For
the second track on April 18, we switched to QSO 0136+478, which
yielded better results ($S_\nu\sim2.8$ Jy). The passband-corrected
spectra, which are featureless, are shown in Figure
\ref{mmspectra}. The sensitivities achieved after passband correction
are $\sigma\sim35$ mJy in the high resolution region. Like most flat
spectrum sources, 2351+456 appears to be variable, with a flux density
of $S_\nu\approx719$ mJy on April 9 and $S_\nu\approx648$ mJy on April 18. 

\section{Results and Analysis}\label{sec:results}

Confusion, coupled with 
RFI-induced effects like slopes and steps in the spectral bandpass, 
indicates that the GBT spectral baseline does not accurately reflect the 
continuum emission from 2351+456.  
Within the 15.5\arcmin\ (FWHM) GBT primary beam at 800 MHz,
there is a contribution of roughly 100 mJy 
at 1.4 GHz from sources of unknown spectral index \citep{con98}.  
We thus interpolate the 
800 MHz flux density from previous 
interferometric measurements:  $1.873\pm 0.056$ Jy at 1.4GHz \citep{con98} 
and $2.158\pm 0.041$ Jy at 365 MHz \citep{dou96} 
indicate a flat-spectrum source
with $\alpha = -0.11\pm 0.03$ ($S_\nu \propto \nu^\alpha$).
Hence, at 798.2257 MHz, the expected (and henceforth assumed) 
flux density is $1.987\pm0.066$ Jy in the absence of variability 
(but see, e.g., \citet{all03}).  

Fits to the \ion{H}{1} profile (Figure \ref{fig:spec}) 
require a minimum of seven Gaussian components
to obtain residuals consistent with the spectral noise.  These fits are not
presented because they are not unique and thus not physical.  
The seven component requirement and the 
width of the absorption system indicates that
the \ion{H}{1} profile is a highly blended, perhaps continuous, ensemble of
gas complexes sampling a galaxy-scale range of velocities and physical 
dimensions.  

The peak optical depth in the \ion{H}{1} line is 0.320 (545.0 mJy) at 
$z = 0.779452 \pm 0.000014$.  We use the most 
conservative error estimate of the redshift of $\pm1$ channel in lieu
of unique reliable Gaussian fits to the profile.  
The integrated optical depth is 13.05 km~s$^{-1}$, and the column 
density is thus 
$N_{HI} = 1.8\times10^{18} (T_s/f) \int \tau\;dv 
    = 2.35\times10^{19} (T_s/f)$ cm$^{-2}$,  where $T_s$ is the
spin temperature in K and $f$ is the covering factor of the continuum 
source by the absorbing gas.   The absorption system qualifies 
as a DLA system ($N_{HI}\geq2\times10^{20}$ cm$^{-2}$)
for $T_s/f\geq8.5$ K.  Typical values span $T_s=100$--1000 K 
\citep{kan03},
indicating
a column density from $N_{HI}=2.35\times10^{21}$ to $2.35\times10^{22}$ 
cm$^{-2}$ for $f=1$.
The FWHM of the absorption line is 53 km s$^{-1}$, and the total span 
of absorption is 163 km s$^{-1}$.  Since the line is unusually broad and
the width is due to bulk velocities rather than to thermal kinetic broadening,
the usual constraint on the kinetic temperature 
$T_k \leq (1.2119\times10^2\ \mbox{FWHM}^2)/(8\ln{2}) \approx 61,000$ K  is 
not physical.

In the BIMA spectra, no HCO$^+$ ($1\rightarrow2$) or HCN ($1\rightarrow2$) 
lines were detected with $|\tau|> 0.15$ (105 mJy) 
at $3\sigma$ significance in 5 km s$^{-1}$ channels (Figure \ref{mmspectra}).
In contrast, the four known molecular absorption systems show large
optical depths, from $\tau=0.32$ in 
HCO$^+$ ($1\rightarrow2$) and $\tau=0.25$ in HCN ($1\rightarrow2$) toward
B3 1504+377 to saturated lines toward PKS 1830$-$211 and B0218+357
\citep{wik95,wik96a,wik96b,wik97}.  The limit on HCO$^+$ absorption 
toward 2351+456
corresponds to $N_{HCO^+} \leq 3.2 \times 10^{13}$ cm$^{-2}$, which translates
into the limit $N_{H_2} \leq 1.6 \times 10^{22}$ cm$^{-2}$ assuming
a gas kinetic temperature of 50 K and 
$N_{H_2} = 5\times10^8\ N_{HCO^+}$ \citep{lis00}.  For $T_k=25$ K, we
obtain $N_{H_2} \leq 5 \times 10^{21}$ cm$^{-2}$, which is equivalent to 
$A_V=2.5$ for a Galactic dust-to-gas ratio.

\section{Discussion} \label{sec:hb89}

The nature of the absorption system toward 2351+456 remains unclear.
The \ion{H}{1} 21 cm absorption profile 
shows striking similarities in width and lack of prominent 
substructure to the absorption toward PKS 1830$-$211 \citep{che99}.  
The absorber toward PKS 1830$-$211 is both a gravitational lens and a 
molecular absorber \citep{sub90,wik96a}, and the unusually broad
and smooth \ion{H}{1} profile likely represents illumination of a significant
fraction of an \ion{H}{1} spiral disk by a background radio source 
magnified by gravitational lensing \citep{pat93,che99}.  
But BIMA observations show that the 2351+456 absorber does not have 
strong molecular absorption lines, and existing VLBI and 
{\it Hubble Space Telescope} ({\it HST}) observations 
can explore the possibility of gravitational lensing of 2351+456.

VLBI observations of 2351+456 at 18 cm (1664 MHz) show a
$\sim20$--30 mas resolved core plus a one-sided jet
morphology with total extent $\sim110$ mas \citep{pol95}.
There is no clear evidence of a second image on scales of tens to 
$\sim150$ mas in the VLBI map (although there is a tantalizingly 
strong knot of emission $\sim110$ mas from the core that could 
be a weak counterimage), and an archival $I$-band (F814W) {\it HST} WFPC2 
image shows no second image on scales $\gtrsim200$ mas.
Although scales relevant to gravitational lensing are
probed by {\it HST} images and VLBI maps from $\sim10$
mas to arcseconds,
no clear companion images have been identified.  We conclude that 2351+456
is unlikely to be gravitationally lensed.

The physical scale subtended by the 18 cm VLBI core at the absorber 
redshift is 0.15--0.22 kpc, but this scale is likely to be larger at 37.5 cm 
(800 MHz).  The absorption line width strongly suggests that the \ion{H}{1}
absorption arises from a unusually large physical scale, of order the size
of the illuminating ``beam'' of the VLBI core.  The high frequency picture
of 2351+456 is dramatically different:  \citet{lister01} shows that 
2351+456 at 43 GHz subtends only $\sim1$ mas.  The millimeter illuminating 
beam is thus only $\sim8$ pc at the absorber redshift, which can potentially
explain the lack of molecular absorption in the BIMA spectra.  Although the
scales probed at centimeter wavelengths are large, the millimeter
scales are in the ``pencil
beam'' regime that makes the detection of molecular absorption so rare
and difficult.

Previous optical observations of 2351+456 may offer some clues to the 
nature of the \hi\ absorption system at $z=0.78$.  
Based on the 1987 September 15 optical spectrum of 
2351+456, which shows a dramatic continuum increase from 
1986 October 26 but preserves flux in emission lines and the equivalent
width in the \ion{Mg}{2} absorption line at $z=1.99$ \citep{law96}, 
we identify additional absorption
lines consistent with two absorbing complexes.  One is roughly at the
redshift of the quasar ($z\approx1.99$), showing possible absorption in 
Ly$\alpha$, 
\ion{Fe}{2} $\lambda\lambda$2344, 2374, 2382, 2586, and 2600,  
\ion{Mg}{2} $\lambda\lambda$2796 and 2803, and 
\ion{Mg}{1} $\lambda$2852 (many of these are blended).  The other is
consistent with the redshift of the \ion{H}{1} detection at $z=0.78$,
with absorption in 
\ion{Fe}{2} $\lambda\lambda$2344 and 2600,
\ion{Mg}{2} $\lambda\lambda$2796 and 2803, and possibly 
H$\gamma$ and H$\delta$ (many of these are blended, and H$\delta$ 
roughly coincides with \ion{Fe}{2} $\lambda$2586 at $z=1.99$, 
so one is likely a spurious identification).  Some of 
these absorption lines can also be seen in the \citet{sti93} spectrum.
\citet{rao00} show that half of absorption 
systems with \ion{Mg}{2} $\lambda$2796 and \ion{Fe}{2} $\lambda$2600
equivalent widths greater than 0.5 \AA\ are DLA systems and that low redshift 
DLA systems that
do not meet these requirements are rare (only one is known).  Hence, although
a higher resolution spectrum is required to extract reliable absorption 
redshifts and equivalent widths, the presence of \ion{Mg}{2} and 
\ion{Fe}{2} absorption lines is consistent with the identification of a
DLA system at $z=0.78$.  

The absorption system toward 2351+456 may also have been mistakenly
identified as the host galaxy of the quasar itself.  
\citet{kuk01} modeled the {\it HST} NICMOS $H$-band surface brightness
of 2351+456 and obtained a half light radius of 18.5 kpc (corrected to 
the concordance cosmology) for the quasar host assuming an $r^{1/4}$ 
model \citep{deV48}.
If the extended emission is in fact from the intervening absorber at $z=0.78$,
then the half light radius of the absorber is 16.2 kpc.  If this is the case, 
then a de Vaucouleurs profile may not be an appropriate parameterized fit.

\section{Conclusions} \label{sec:conclusions}

We have identified a DLA system toward 2351+456
via the \ion{H}{1} 21 cm absorption line.
We identify \ion{Mg}{2} $\lambda\lambda$2796 and 2803 and 
\ion{Fe}{2} $\lambda\lambda$2344
and 2600 lines in the optical spectrum of \citet{law96}, in support of the
\ion{H}{1} 21 cm absorption line at $z=0.78$ being a DLA system.  
Although the \ion{H}{1} absorption line profile of 2351+456 is remarkably 
similar to the profile of PKS 1830$-$211, the DLA system toward 2351+456
does not appear to be either a 
molecular absorption system or a gravitational lens.  
This new detection
provides a proof-of-concept for blind \ion{H}{1} surveys with broadband 
radio receivers and backends that offer a new dust- and atmosphere-independent
method to identify DLA systems.  Broadband radio surveys for DLA systems
are perhaps key to identifying exceptionally rare molecular absorption systems
and may identify gravitational lenses as well.

\acknowledgments

We are grateful to the staff of the Green Bank Telescope
for observing and data reduction support, especially Glen Langston,
Bob Garwood, and Carl Bignell.
R. G. and M. P. H. have received partial support from NSF grant AST-0307661.
This research has made use of the NASA/IPAC Extragalactic Database,
which is operated by the Jet Propulsion Laboratory, California Institute 
of Technology, under contract with the National Aeronautics and Space 
Administration.
Some of the data presented in this paper were obtained from the 
Multimission Archive at the Space Telescope Science Institute. 
STScI is operated by the Association of Universities for Research in 
Astronomy, Inc., under NASA contract NAS5-26555. 



\clearpage



\end{document}